\documentclass[12pt]{amsart}
\usepackage{amssymb,hyperref}

\newtheorem{theorem}{Theorem}[section]
\newtheorem{definition}[theorem]{Definition}
\newtheorem{conjecture}[theorem]{Conjecture}
\newtheorem{proposition}[theorem]{Proposition}
\newtheorem{corollary}[theorem]{Corollary}

\newtheorem{lemma}[theorem]{Lemma}

\newcommand{\ba}{{\bf a}}
\newcommand{\bb}{{\bf b}}
\newcommand{\bc}{{\bf c}}
\newcommand{\bd}{{\bf d}}

\begin{document}

\title{On polynomial integrals over the orthogonal group}

\author{Teodor Banica}
\address{T.B.: Department of Mathematics, Cergy-Pontoise University, 95000 Cergy-Pontoise, France. {\tt teodor.banica@u-cergy.fr}}

\author{Benoit Collins}
\address{B.C.: Department of Mathematics, Lyon 1 University, and University of Ottawa, 585 King Edward, Ottawa, ON K1N 6N5, Canada. {\tt bcollins@uottawa.ca}}

\author{Jean-Marc Schlenker}
\address{J.-M.S.: Institut de Math\'ematiques de Toulouse, 
UMR CNRS 5219, Universit\'e Toulouse 3, 118 route de Narbonne, 
31062 Toulouse Cedex 9, France. {\tt schlenker@math.univ-toulouse.fr}}


\begin{abstract}
We consider integrals of type $\int_{O_n}u_{11}^{a_1}\ldots u_{1n}^{a_n}u_{21}^{b_1}\ldots u_{2n}^{b_n}\,du$, with respect to the Haar measure on the orthogonal group. We establish several remarkable invariance properties satisfied by such integrals, by using combinatorial methods. We present as well a general formula for such integrals, as a sum of products of factorials.
\end{abstract}

\maketitle

\section*{Introduction}

The computation of polynomial integrals over the orthogonal group $O_n$ is a key problem in mathematical physics. These integrals are indeed known to appear in a wealth of concrete situations, coming from random matrices, lattice models, combinatorics.

These integrals are best introduced in a ``rectangular way'', as follows:
$$I(a)=\int_{O_n}\prod_{i=1}^p\prod_{j=1}^qu_{ij}^{a_{ij}}\,du$$

For some previous work on the subject, mostly of asymptotic nature, see \cite{ban}, \cite{cma}, \cite{csn}, \cite{dsh}, \cite{mno}, \cite{ner}, \cite{nov}, \cite{ols}, \cite{pre}, \cite{wei}, \cite{zin}. For noncommutative versions, see \cite{bcz}, \cite{bgo}, \cite{cur}. For a motivation for the exact computation of $I(a)$, coming from Hadamard matrices, see \cite{bcs}.

In the one-row case the above integral takes place over the ``first slice'' of $O_n$, known to be isomorphic to the sphere $S^{n-1}$, and a standard computation gives:
$$I\begin{pmatrix}a_1&\ldots&a_q\end{pmatrix}=\frac{(n-1)!!a_1!!\ldots a_q!!}{(\Sigma a_i+n-1)!!}$$

In this paper we investigate the two-row case. More precisely, we are interested in the exact computation of the following integrals, depending on $a_i,b_i\in 2\mathbb N$:
$$I\begin{pmatrix}a_1&\ldots&a_q\\ b_1&\ldots&b_q\end{pmatrix}=\int_{O_n}u_{11}^{a_1}\ldots u_{1q}^{a_q}u_{21}^{b_1}\ldots u_{2q}^{b_q}\,du$$
\b
It is convenient to consider the vectors  $\ba,\bb$ with coordinates $a_i, b_j$ respectively -- 
in this paper we will use bold letters to denote vectors or matrices.   
We make the following normalization, where $I_{n-1}$ denotes the polynomial integration in the above sense, over the group $O_{n-1}$:
$$I\begin{pmatrix}\ba\\ \bb\end{pmatrix}=I_{n-1}\begin{pmatrix}\ba\end{pmatrix}I_{n-1}\begin{pmatrix}\bb\end{pmatrix}\Phi\begin{pmatrix}\ba\\ \bb\end{pmatrix}$$

The point with this normalization is that the $\Phi$ quantity on the right has a number of remarkable symmetry properties. We have the following result:

\medskip

\noindent {\bf Theorem A.} {\em The function $\Phi=\Phi_n$ has the following properties:
\begin{enumerate}
\item Flipping: $\Phi(^\ba_\bb{\ }^\bc_\bd)=\Phi(^\ba_\bb{\ }^\bd_\bc)$.

\item Compression: $\Phi(^\ba_\bb{\ }^\bc_0)=\Phi(^\ba_\bb{\ }^{\Sigma c_i}_{\ 0})$.

\item Transmutation: $\Phi_n(^\ba_\bb{\ }^2_0)=(1-1/n)\Phi_{n+2}(^\ba_\bb)$.
\end{enumerate}}

\medskip

In this statement the main result is the first one. Quite curiously, this simple formula seems to resist any kind of direct geometric approach, or conceptual understanding in general. In what follows we will present a heavily combinatorial proof for it.

Our second result is an exact formula for $\Phi$, as a sum of products of factorials.

\medskip

\noindent {\bf Theorem B.} {\em The values of $\Phi$ are given by
$$\Phi\begin{pmatrix}2\ba\\ 2\bb\end{pmatrix}=\frac{(n-1)!!}{(n-2)!!}\sum_{r_1\ldots r_q}(-1)^R\prod_{i=1}^q\frac{4^{r_i}a_i!b_i!}{(2r_i)!(a_i-r_i)!(b_i-r_i)!}\cdot\frac{(2R)!!(2S-2R+n-2)!!}{(2S+n-1)!!}$$
where the sum is over $r_i=0,1,\ldots,\min(a_i,b_i)$, and $S=\Sigma a_i+\Sigma b_i,R=\Sigma r_i$.}
\medskip

As an illustration, let us look at the simplest case, when only one entry of $a$ and of $b$ is nonzero. After doing some standard manipulations, we obtain the following result.

\medskip

\noindent {\bf Theorem C.} {\em The joint moments of $2$ orthogonal group coordinates $x,y\in\{u_{ij}\}$, chosen in generic position (i.e. not on the same row or column), are given by
$$\int_{O_n}x^\alpha y^\beta\,du=\frac{(n-2)!\alpha!!\beta!!(\alpha+\beta+n-2)!!}{(\alpha+n-2)!!(\beta+n-2)!!(\alpha+\beta+n-1)!!}$$
for $\alpha,\beta$ even, and vanish if one of $\alpha,\beta$ is odd.}

\medskip

Let us point out the fact that this kind of technical formula is quite powerful. For instance the $n\to\infty$ behavior of the above quantity can be simply obtained by using the Stirling formula, and we recover in this way a well-known result of Diaconis and Shahshahani \cite{dsh}, stating that with $n\to\infty$ the variables $x,y$ are Gaussian and independent.

Of course, there are several other interesting formulae emerging from Theorem A and Theorem B. We will explore them, with full explanations, in the body of the paper.

The present paper is purely computational. For potential applications we refer to the papers cited in the reference list, and to the papers cited in that papers. 

Let us mention here, however, that we have in mind two kinds of applications. With the notation $k=\Sigma a_{ij}$, these potential applications fall into two classes, as follows:

\begin{enumerate}
\item Case ``$k$ fixed, $n\to\infty$''. This is somehow the ``old'' problematics, coming from probability, random matrices, lattice models. As already explained, our exact results can lead to asymptotic ones simply by using the Stirling formula.

\item Case ``$n$ fixed, $k\to\infty$''. Here the idea, needed for instance in relation with some difficult problems in combinatorics \cite{bcs}, would be to obtain, once again via the Stirling formula, estimates at $n$ fixed, for classes of matrices with $\Sigma a_{ij}\to\infty$.
\end{enumerate}

Now back to the general problem of computing $I(\ba)$, it is our hope that the present results might serve as a ``key input'' for this purpose. Indeed, most of our statements seem to have some natural generalizations to the multi-row case, and the problem is of course to check the validity of these generalized statements. We intend to investigate the $p$-row case, with $p=3$, or perhaps even bigger, in a future paper.

The paper is organized as follows: 1-2 are preliminary sections, in 3 we present an alternative approach to the Weingarten formula, specially designed for the two-row case, and in 4-6 we state and prove the main results. The final sections, 7-8, contain a number of consequences and refinements in the case $\ba\in M_2(\mathbb N)$, and a few concluding remarks.

\subsection*{Acknowledgements}

We would like to thank J. Novak for several useful discussions. T.B. and B.C. were supported by the ANR grants ``Galoisint'' and ``Granma''. B.C. was also supported by NSERC.

\section{Group integrals}

The polynomial integrals over the orthogonal group $O_n$, our basic object of study, are best introduced in a ``rectangular form'', as follows.

\begin{definition}
Associated to any matrix $\ba\in M_{p\times q}(\mathbb N)$ is the integral
$$I(\ba)=\int_{O_n}\prod_{i=1}^p\prod_{j=1}^qu_{ij}^{a_{ij}}\,du$$
with respect to the uniform measure on the orthogonal group $O_n$.
\end{definition}

In this definition, and also in most of the statements to follow, we prefer to be a bit unprecise on the meaning of the variable $n$. Normally in the above situation $n$ should be an integer greater than $p,q$; but the results in section 2 below show that the integration over $O_n$ can be given a purely formal meaning, so that $n$ can be any kind of variable.

For some previous work on the subject, of asymptotic nature, see \cite{ban}, \cite{cma}, \cite{csn}, \cite{dsh}, \cite{mno}, \cite{ner}, \cite{ols}, \cite{pre}, \cite{wei}, \cite{zin}. For a motivation for the exact computation of $I(a)$, see \cite{bcs}.

The advantage of our rectangular formulation comes from the fact that the parameters $p,q\in\mathbb N$ effectively measure the ``complexity'' of the computation. For instance  in the one-row case ($p=1$), we have the following elementary, well-known result.

\begin{theorem}
For any $a_1,\ldots,a_q$ even we have the formula
$$I\begin{pmatrix}a_1&\ldots&a_q\end{pmatrix}=\frac{(n-1)!!a_1!!\ldots a_q!!}{(\Sigma a_i+n-1)!!}$$
where $m!!=(m-1)(m-3)(m-5)\ldots$, with the product ending at $1$ or $2$.
\end{theorem}

\begin{proof}
This follows from the well-known fact that the first slice of $O_n$ is isomorphic to the real sphere $S^{n-1}$. Indeed, this gives the following formula:
$$I\begin{pmatrix}a_1&\ldots&a_q\end{pmatrix}=\int_{S^{n-1}}x_1^{a_1}\ldots x_q^{a_q}\,dx$$ 

The integral on the right can be computed by using polar coordinates and the Fubini theorem, and we obtain the formula in the statement. See e.g. \cite{bcs}. 
\end{proof}

Another well-known result, of trigonometric nature as well, is as follows.

\begin{theorem}
At $n=2$ we have the formula
$$I\begin{pmatrix}a&b\\ c&d\end{pmatrix}=\varepsilon\cdot\frac{(a+d)!!(b+c)!!}{(a+b+c+d+1)!!}$$
where $\varepsilon=1$ if $a,b,c,d$ are even, $\varepsilon=-1$ is $a,b,c,d$ are odd, and $\varepsilon=0$ otherwise.
\end{theorem}

\begin{proof}
When computing the integral over $O_2$, we can restrict the integration to $SO_2=S^1$, then further restrict the integration to the first quadrant. We get: 
$$I\begin{pmatrix}a&b\\ c&d\end{pmatrix}
=\varepsilon\cdot\frac{2}{\pi}\int_0^{\pi/2}(\cos t)^{a+d}(\sin t)^{b+c}\,dt$$

This gives the formula in the statement.
\end{proof}

Finally, a third elementary result about $I(\ba)$ is as follows.

\begin{theorem}
The integral $I(\ba)$ vanishes unless the matrix $\ba\in M_{p\times q}(\mathbb N)$ is ``admissible'', in the sense that the sum on each of its rows and columns is an even number.
\end{theorem}

\begin{proof}
This follows by multiplying the rows or columns of $u$ by $-1$, and by using the basic invariance properties of the Haar measure on $O_n$.
\end{proof}

Observe in particular that in the $2\times 2$ case, the admissible matrices are those having all entries even, or all entries odd. This agrees of course with Theorem 1.3.

\section{The Weingarten formula}

Our main tool for the computation of integrals over $O_n$ will be a combinatorial formula, whose origins go back to Weingarten's paper \cite{wei}. In this section we make a brief presentation of the formula, as developed in \cite{csn}, and then we present a combinatorial interpretation of the Weingarten matrix entries, to be heavily used in what follows.

Given a pairing $\pi$ and a multi-index $i$ we say that ``$i$ fits into $\pi$'' if, when putting the indices of $i$ on the points of $\pi$,  each string of $\pi$ connects a pair of equal indices.

\begin{theorem}
We have the Weingarten formula
$$\int_{O_n}u_{i_1j_1}\ldots u_{i_{2k}j_{2k}}\,du=\sum_{\pi,\sigma\in D_k}\delta_\pi(i)\delta_\sigma(j)W_{kn}(\pi,\sigma)$$
where the objects on the right are as follows:
\begin{enumerate}
\item $D_k$ is the set of pairings of $\{1,\ldots,2k\}$.

\item The delta symbols are $1$ or $0$, depending on whether indices fit or not.

\item The Weingarten matrix is $W_{kn}=G_{kn}^{-1}$, where $G_{kn}(\pi,\sigma)=n^{|\pi\vee\sigma|}$.
\end{enumerate}
\end{theorem}

\begin{proof}
The idea is that integrals on the left form the orthogonal projection onto $Fix(u^{\otimes 2k})$, which is spanned by the vectors $\xi_\pi=\Sigma_i\delta_\pi(i)e_{i_1}\otimes\ldots\otimes e_{i_{2k}}$. Since the Gram matrix of these vectors is $<\xi_\pi,\xi_\sigma>=G_{kn}(\pi,\sigma)$, we obtain the formula in the statement. See \cite{csn}.
\end{proof}

As an example, the integrals of quantities of type $u_{i_1j_1}u_{i_2j_2}u_{i_3j_3}u_{i_4j_4}$ appear as sums of coefficients of the Weingarten matrix $W_{2n}$, which is given by:
$$W_{2n}=\begin{pmatrix}
n^2&n&n\\ 
n&n^2&n\\ 
n&n&n^2\end{pmatrix}^{-1}
=\frac{1}{n(n-1)(n+2)}
\begin{pmatrix}
n+1&-1&-1\\ 
-1&n+1&-1\\ 
-1&-1&n+1\end{pmatrix}$$

More precisely, the various consequences at $k=2$ can be summarized as follows.

\begin{proposition}
We have the following results:
\begin{enumerate}
\item $I(^4_0{\ }^0_0)=3/(n(n+2))$.

\item $I(^2_0{\ }^2_0)=1/(n(n+2))$.

\item $I(^2_0{\ }^0_2)=(n+1)/(n(n-1)(n+2))$.
\end{enumerate}
\end{proposition}

\begin{proof}
These results all follow from the Weingarten formula, by using the above numeric values for the entries of $W_{2n}$:
\begin{eqnarray*}
I\begin{pmatrix}4&0\\ 0&0\end{pmatrix}
&=&\int u_{11}u_{11}u_{11}u_{11}=\sum_{\pi\sigma}W_{2n}(\pi,\sigma)=\frac{3(n+1)-6}{n(n-1)(n+2)}=\frac{3}{n(n+2)}\\
I\begin{pmatrix}2&2\\ 0&0\end{pmatrix}
&=&\int u_{11}u_{11}u_{12}u_{12}=\sum_\pi W_{2n}(\pi,\cap\cap)=\frac{(n+1)-2}{n(n-1)(n+2)}=\frac{1}{n(n+2)}\\
I\begin{pmatrix}2&0\\ 0&2\end{pmatrix}
&=&\int u_{11}u_{11}u_{22}u_{22}=W_{2n}(\cap\cap,\cap\cap)=\frac{n+1}{n(n-1)(n+2)}
\end{eqnarray*}

Here $\cap\cap$ denotes the pairing of $\{1,2,3,4\}$ which pairs 1 with 2, and 3 with 4.

Observe that the first and second formulae follow in fact as well from Theorem 1.2.
\end{proof}

In general, the computation of the Weingarten matrix is a quite subtle combinatorial problem, and the first results here go back to \cite{wei}, \cite{csn}. A quite powerful formula, which is however not exactly adapted to the ``symmetry searching'' considerations in this paper, was recently obtained in \cite{cma}, and was further processed and clarified in \cite{zin}.

The interpretation of the Weingarten matrix that we will need here is in terms of the 0-1-2 matrices having sum 2 on each column. We call such matrices ``elementary''.

\begin{theorem}
The Weingarten matrix entries are given by
$$W_{kn}(\pi,\sigma)=I(\ba)$$
where $\ba\in M_k(\mathbb N)$ is the elementary matrix obtained as follows:
\begin{enumerate}
\item Label $\pi_1,\ldots,\pi_k$ the strings of $\pi$.

\item Label $\sigma_1,\ldots,\sigma_k$ the strings of $\sigma$.

\item Set $a_{ij}=\#\{r\in \{1,\ldots,2k\}|r\in\pi_i,r\in\sigma_j\}$.
\end{enumerate}
\end{theorem}

\begin{proof}
Consider the multi-indices $i,j\in\{1,\ldots,k\}^{2k}$ given by $i_r\in\pi_r$ and $j_r\in\sigma_r$, for any $r\in\{1,\ldots,k\}$. We have $\delta_{\pi'}(i)=\delta_{\pi\pi'}$ and $\delta_{\sigma'}(j)=\delta_{\sigma\sigma'}$ for any pairings $\pi',\sigma'$, so if we apply the Weingarten formula to the quantity $u_{i_1j_1}\ldots u_{i_{2k}j_{2k}}$, we obtain:
\begin{eqnarray*}
\int_{O_n}u_{i_1j_1}\ldots u_{i_{2k}j_{2k}}\,du
&=&\sum_{\pi'\sigma'}\delta_{\pi'}(i)\delta_{\sigma'}(j)W_{kn}(\pi',\sigma')\\
&=&\sum_{\pi'\sigma'}\delta_{\pi\pi'}\delta_{\sigma\sigma'}W_{kn}(\pi',\sigma')\\
&=&W_{kn}(\pi,\sigma)
\end{eqnarray*}

The integral on the left can be written in the form $I(\ba)$, for a certain matrix $\ba$. Our choice of $i,j$ shows that $\ba$ is the elementary matrix in the statement, and we are done.
\end{proof}

As an illustration for the above result, consider the partitions $\pi=\cap\cap\cap$ and $\sigma=\Cap\cap$. We have $i=(112233)$ and $j=(122133)$, and we obtain:
\begin{eqnarray*}
W_{3n}(\pi,\sigma)
&=&\int_{O_n}u_{11}u_{12}u_{22}u_{21}u_{33}u_{33}\,du\\
&=&\int_{O_n}u_{11}u_{12}u_{22}u_{21}u_{33}^2\,du\\
&=&I\begin{pmatrix}1&1&0\\ 1&1&0\\ 0&0&2\end{pmatrix}
\end{eqnarray*}

As a first application of the above Weingarten methods, let us discuss the $n\to\infty$ behavior of $I(\ba)$. Note first that with $n\to\infty$ the above matrix $W_{2n}$ is concentrated on the diagonal. This property holds in fact for any $k$, and has the following consequence.

\begin{theorem}
With $k=\Sigma a_{ij}$ we have the estimate
$$I(\ba)=n^{-k}\left(\prod_{i=1}^p\prod_{j=1}^qa_{ij}!!+O(n^{-1})\right)$$
when all $a_{ij}$ are even, and $I(\ba)=O(n^{-k-1})$ if not.
\end{theorem}

\begin{proof}
This result, known from \cite{dsh}, follows from Theorem 2.1, because with $n\to\infty$ the matrix $G_{kn}$, and hence the matrix $W_{kn}$ too, is concentrated on the diagonal. See \cite{ban}.
\end{proof}

We can see from the above result, and also from the explicit formulae given in section 1, that $I(\ba)$ depends in a non-trivial way on the parity of the entries $a_{ij}$. 

In what follows we will basically focus on the ``main case'', where all entries $a_{ij}$ are even numbers. The general case will be discussed at the end of section 7 below.

\section{Elementary expansion}

In this section and in the next few ones we investigate the integrals of type $I(\ba)$, in the 2-row case. It is convenient to make the following normalization, where, as usual, the general formality comments given in the beginning of section 1 apply.

\begin{definition}
For $\ba,\bb$ vectors with even entries we make the normalization
$$I\begin{pmatrix}\ba\\ \bb\end{pmatrix}=I_{n-1}\begin{pmatrix}\ba\end{pmatrix}I_{n-1}\begin{pmatrix}\bb\end{pmatrix}\Phi\begin{pmatrix}\ba\\ \bb\end{pmatrix}$$
where $I_{n-1}$ denotes the integration in the sense of Definition 1.1, over the group $O_{n-1}$.
\end{definition}

This new quantity $\Phi$ is just a normalization of the usual integral $I$. More precisely, by using the formula in Theorem 1.2 we have the following alternative definition.

\begin{proposition}
We have the formula:
$$\Phi\begin{pmatrix}\ba\\ \bb\end{pmatrix}=\frac{(\Sigma a_i+n-2)!!(\Sigma b_i+n-2)!!}{(n-2)!!(n-2)!!\prod a_i!!\prod b_i!!}\,I\begin{pmatrix}\ba\\ \bb\end{pmatrix}$$
\end{proposition}

\begin{proof}
This follows indeed from the one-row formula in Theorem 1.2.
\end{proof}

As a first, basic example, for any one-row vector $\ba$ we have $\Phi(^\ba_0)=I_n(\ba)/I_{n-1}(\ba)$, and according to Theorem 1.2, this gives the following formula:
$$\Phi\begin{pmatrix}\ba\\ 0\end{pmatrix}=\frac{(n-1)!!}{(n-2)!!}\cdot\frac{(\Sigma a_i+n-2)!!}{(\Sigma a_i+n-1)!!}$$

The advantage of using $\Phi$ instead of $I$ comes from a number of remarkable invariance properties at the general level, to be established later on.

For $k,x\in\mathbb N$ we let $k^x=k\ldots k$ ($x$ times). With this notation, we have the following technical version of Theorem 2.3, to be heavily used in what follows.

\begin{theorem}
We have the ``elementary expansion'' formula
$$\Phi\begin{pmatrix}2\ba\\ 2\bb\end{pmatrix}=\sum_{r_1\ldots r_q}\prod_{i=1}^q\frac{4^{r_i}a_i!b_i!}{(2r_i)!(a_i-r_i)!(b_i-r_i)!}\Phi\begin{pmatrix}1^{2R}&2^{A-R}&0^{B-R}\\
1^{2R}&0^{A-R}&2^{B-R}
\end{pmatrix}$$
where the sum is over $r_i=0,1,\ldots,\min(a_i,b_i)$, and $A=\Sigma a_i,B=\Sigma b_i,R=\Sigma r_i$.
\end{theorem}

\begin{proof}
We use the same method as in the proof of Theorem 2.3. Let us first apply the Weingarten formula to the integral in the statement: 
\begin{eqnarray*}
I\begin{pmatrix}2\ba\\ 2\bb\end{pmatrix}
&=&\int_{O_n}u_{11}^{2a_1}\ldots u_{1q}^{2a_q}u_{21}^{2b_1}\ldots u_{2q}^{2b_q}\,du\\
&=&\sum_{\pi\sigma}\delta_\pi(1^{2A}2^{2B})\delta_\sigma(1^{2a_1}\ldots q^{2a_q}1^{2b_1}\ldots q^{2b_q})W_{kn}(\pi,\sigma)\\
&=&\sum_\sigma \delta_\sigma(1^{2a_1}\ldots q^{2a_q}1^{2b_1}\ldots q^{2b_q})\sum_\pi\delta_\pi(1^{2A}2^{2B})W_{kn}(\pi,\sigma)
\end{eqnarray*}

Now let us look at $\sigma$. In order for the $\delta_\sigma$ symbol not to vanish, $\sigma$ must connect between themselves the $2a_1+2b_1$ copies of $1$, the $2a_2+2b_2$ copies of 2, and so on, up to the $2a_q+2b_q$ copies of $q$. So, for any $i\in\{1,\ldots,q\}$, let us denote by $2r_i\in\{0,2,\ldots,\min(2a_i,2b_i)\}$ the number of ``type $a$'' copies of $i$ coupled with ``type $b$'' copies of $i$.

Our claim is that when these parameters $r_1,\ldots,r_q$ are fixed, the sum on the right doesn't depend on $\sigma$, and provides us with a decomposition of the following type:
$$I\begin{pmatrix}2\ba\\ 2\bb\end{pmatrix}=\sum_{r_1\ldots r_q}N_r(\ba,\bb)I_r(\ba,\bb)$$

Indeed, let us label $\sigma_1,\ldots,\sigma_k$ the strings of $\sigma$, and consider the multi-index $j\in\{1,\ldots,k\}^{2k}$ given by $j_r\in\sigma_r$, for any $r\in\{1,\ldots,k\}$. We have $\delta_{\sigma'}(j)=\delta_{\sigma\sigma'}$ for any pairing $\sigma'$, so by applying once again the Weingarten formula we obtain:
\begin{eqnarray*}
\int_{O_n}u_{1j_1}\ldots u_{1j_{2A}}u_{2j_{2A+1}}\ldots u_{2j_{2A+2B}}\,du
&=&\sum_{\pi\sigma'}\delta_\pi(1^{2A}2^{2B})\delta_{\sigma'}(j)W_{kn}(\pi,\sigma')\\
&=&\sum_{\pi\sigma'}\delta_\pi(1^{2A}2^{2B})\delta_{\sigma\sigma'}W_{kn}(\pi,\sigma')\\
&=&\sum_\pi\delta_\pi(1^{2A}2^{2B})W_{kn}(\pi,\sigma)
\end{eqnarray*}

Now let us look at the integral on the left. This can be written in the form $I(m)$, for a certain matrix $m$, the procedure being simply to group together, by using exponents, the identical terms in the product of $u_{ij}$'s. Now by getting back to the definition of the multi-index $j$, we can conclude that this procedure leads to the following formula:
$$\int_{O_n}u_{1j_1}\ldots u_{1j_{2A}}u_{2j_{2A+1}}\ldots u_{2j_{2A+2B}}\,du
=I\begin{pmatrix}1^{2R}&2^{A-R}&0^{B-R}\\
1^{2R}&0^{A-R}&2^{B-R}\end{pmatrix}$$

Summing up, our claim is proved, and the quantity $I_r(\ba,\bb)$ is nothing but the integral in the statement. That is, we have proved the following formula, where $N_r(\ba,\bb)$ is the number of pairings $\sigma$ as those considered above:
$$I\begin{pmatrix}2\ba\\ 2\bb\end{pmatrix}=\sum_{r_1\ldots r_q}N_r(\ba,\bb)I\begin{pmatrix}1^{2R}&2^{A-R}&0^{B-R}\\
1^{2R}&0^{A-R}&2^{B-R}\end{pmatrix}$$

Let us compute now the coefficient $N_r(\ba,\bb)$. This is by definition the number of pairings $\sigma$ as above, and these pairings are obtained as follows: (1) pick $2r_i$ elements among $2a_i$ elements, (2) pick $2r_i$ elements among $2b_i$ elements, (3) couple the ``type $a$'' $2r_i$ elements to the ``type $b$'' $2r_i$ elements, (4) couple the remaining $2a_i-2r_i$ elements, (5) couple the remaining $2b_i-2r_i$ elements. Thus we have:
\begin{eqnarray*}
N_r(\ba,\bb)
&=&\prod_{i=1}^q\begin{pmatrix}2a_i\\ 2r_i\end{pmatrix}\begin{pmatrix}2b_i\\ 2r_i\end{pmatrix}(2r_i)!(2a_i-2r_i)!!(2b_i-2r_i)!!\\
&=&\prod_{i=1}^q\frac{(2a_i)!(2b_i)!(2r_i)!(2a_i-2r_i)!!(2b_i-2r_i)!!}{(2r_i)!(2a_i-2r_i)!(2r_i)!(2b_i-2r_i)!}\\
&=&\prod_{i=1}^q\frac{(2a_i)!(2b_i)!}{(2r_i)!(2a_i-2r_i+1)!!(2b_i-2r_i+1)!!}
\end{eqnarray*}

Summing up, we have proved the following formula:
$$I\begin{pmatrix}2\ba\\ 2\bb\end{pmatrix}=\sum_{r_1\ldots r_q}\prod_{i=1}^q\frac{(2a_i)!(2b_i)!}{(2r_i)!(2a_i-2r_i+1)!!(2b_i-2r_i+1)!!}I\begin{pmatrix}1^{2R}&2^{A-R}&0^{B-R}\\ 1^{2R}&0^{A-R}&2^{B-R}\end{pmatrix}$$

It remains to convert this formula in terms of $\Phi$ quantities. But this can be done by using Proposition 3.2. By applying it twice, we get:
$$\Phi\begin{pmatrix}2\ba\\ 2\bb\end{pmatrix}
=\frac{(2A+n-2)!!(2B+n-2)!!}{(n-2)!!(n-2)!!\prod (2a_i)!!\prod (2b_i)!!}\,I\begin{pmatrix}2\ba\\ 2\bb\end{pmatrix}$$
$$\Phi\begin{pmatrix}1^{2R}&2^{A-R}&0^{B-R}\\ 1^{2R}&0^{A-R}&2^{B-R}\end{pmatrix}
=\frac{(2A+n-2)!!(2B+n-2)!!}{(n-2)!!(n-2)!!}\,I\begin{pmatrix}1^{2R}&2^{A-R}&0^{B-R}\\ 1^{2R}&0^{A-R}&2^{B-R}\end{pmatrix}$$

Thus when passing to $\Phi$ quantities, the only thing that happens is that the numeric coefficient gets divided by $\prod (2a_i)!!\prod (2b_i)!!$. So, this coefficient becomes:
\begin{eqnarray*}
N_r'(\ba,\bb)
&=&\prod_{i=1}^q\frac{1}{(2a_i)!!(2b_i)!!}\prod_{i=1}^q\frac{(2a_i)!(2b_i)!}{(2r_i)!(2a_i-2r_i+1)!!(2b_i-2r_i+1)!!}\\
&=&\prod_{i=1}^q\frac{(2a_i+1)!!(2b_i+1)!!}{(2r_i)!(2a_i-2r_i+1)!!(2b_i-2r_i+1)!!}\\
&=&\prod_{i=1}^q\frac{4^{r_i}a_i!b_i!}{(2r_i)!(a_i-r_i)!(b_i-r_i)!}
\end{eqnarray*}

Thus we have obtained the formula in the statement, and we are done.
\end{proof}

As a first consequence, we have the following result.

\begin{theorem}
We have the ``compression formula''
$$\Phi\begin{pmatrix}\ba&\bc\\ \bb&0\end{pmatrix}
=\Phi\begin{pmatrix}\ba&\Sigma c_i\\ \bb&0\end{pmatrix}$$
valid for any vectors with even entries $\ba,\bb\in\mathbb N^p$ and $\bc\in\mathbb N^q$.
\end{theorem}

\begin{proof}
It is convenient to replace $\ba,\bb,\bc$ with their doubles $2\ba,2\bb,2\bc$. Consider now the elementary expansion formula for the matrix in the statement:
$$\Phi\begin{pmatrix}2\ba&2\bc\\ 2\bb&0\end{pmatrix}=\sum_{r_1\ldots r_q}\prod_{i=1}^q\frac{4^{r_i}a_i!b_i!}{(2r_i)!(a_i-r_i)!(b_i-r_i)!}\Phi\begin{pmatrix}1^{2R}&2^{A+C-R}&0^{B-R}\\
1^{2R}&0^{A+C-R}&2^{B-R}
\end{pmatrix}$$

Since the numeric coefficient doesn't depend on $c$, and the function on the right depends only on $C=\Sigma c_i$, this gives the formula in the statement.
\end{proof}

We should mention that the above formula has as well a direct geometric proof. However, this is no longer true for the ``flipping formula'' in section 5 below, or for the ``transmutation formula'' in section 6 below. So, as a general policy, in what follows we will simply present combinatorial proofs for all the results. The geometric methods that we have so far are not powerful enough, and will be rather explained in some future paper.

\section{Triangular formula}

We explore now a problematics which is somehow opposite to the ``compression principle'': what happens when ``extending'' the original matrix $(^\ba_\bb)$ with a $(^\bc_0)$ component?

In this section we present a number of technical results in this sense, which will lead to a number of concrete formulae, of great use for the general purposes of this paper. The final answer to the ``extension problem'' will be given later on (Theorem 6.4 below).

Let us begin with a basic lemma.

\begin{lemma}
We have the ``basic extension'' formula
$$\Phi\begin{pmatrix}\ba&2\\ \bb&0\end{pmatrix}
=\frac{1}{n-q}\left((\Sigma a_i+n-1)\Phi\begin{pmatrix}\ba\\ \bb\end{pmatrix}-\sum_{s=1}^q(a_s+1)\Phi\begin{pmatrix}\ba^{(s)}\\ \bb\end{pmatrix}\right)$$
for any $\ba,\bb\in(2\mathbb N)^q$, where $\ba^{(s)}=(a_1,\ldots,a_{s-1},a_s+2,a_{s+1},\ldots,a_q)$.
\end{lemma}

\begin{proof}
By using the trivial identity $\Sigma u_{1i}^2=1$, we obtain the following formula:
$$I\begin{pmatrix}\ba\\ \bb\end{pmatrix}=\sum_{s=1}^qI\begin{pmatrix}\ba^{(s)}\\ \bb\end{pmatrix}+(n-q)I\begin{pmatrix}\ba&2\\ \bb&0\end{pmatrix}$$

Let us translate this formula in terms of $\Phi$ quantities. According to Proposition 3.2, we have:
$$\Phi\begin{pmatrix}\ba\\ \bb\end{pmatrix}=\frac{(\Sigma a_i+n-2)!!(\Sigma b_i+n-2)!!}{(n-2)!!(n-2)!!\prod a_i!!\prod b_i!!}\,I\begin{pmatrix}\ba\\ \bb\end{pmatrix}$$
$$\Phi\begin{pmatrix}\ba^{(s)}\\ \bb\end{pmatrix}=\frac{(\Sigma a_i+n)!!(\Sigma b_i+n-2)!!}{(n-2)!!(n-2)!!\prod a_i!!\prod b_i!!(a_s+1)}\,I\begin{pmatrix}\ba^{(s)}\\ \bb\end{pmatrix}$$
$$\Phi\begin{pmatrix}\ba&2\\ \bb&0\end{pmatrix}=\frac{(\Sigma a_i+n)!!(\Sigma b_i+n-2)!!}{(n-2)!!(n-2)!!\prod a_i!!\prod b_i!!}\,I\begin{pmatrix}\ba&2\\ \bb&0\end{pmatrix}$$

Thus our above formula translates as follows:
$$(\Sigma a_i+n-1)\Phi\begin{pmatrix}\ba\\ \bb\end{pmatrix}=\sum_{s=1}^q(a_s+1)\Phi\begin{pmatrix}\ba^{(s)}\\ \bb\end{pmatrix}+(n-q)\Phi\begin{pmatrix}\ba&2\\ \bb&0\end{pmatrix}$$

This gives the formula in the statement.
\end{proof}

It is convenient to record as well a ``recursive'' version of the above result.

\begin{lemma}
We have the ``recursive extension'' formula
$$\Phi\begin{pmatrix}\ba&c+2\\ \bb&0\end{pmatrix}
=\frac{1}{n+c-q}\left((\Sigma a_i+c+n-1)\Phi\begin{pmatrix}\ba&c\\ \bb&0\end{pmatrix}-\sum_{s=1}^q(a_s+1)\Phi\begin{pmatrix}\ba^{(s)}&c\\ \bb&0\end{pmatrix}\right)$$
valid for any two vectors $\ba,\bb\in(2\mathbb N)^q$, and any $c\in 2\mathbb N$.
\end{lemma}

\begin{proof}
We use the compression formula. This gives:
$$\Phi\begin{pmatrix}\ba&c+2\\ \bb&0\end{pmatrix}=\Phi\begin{pmatrix}\ba&c&2\\ \bb&0&0\end{pmatrix}$$

Now if we denote the quantity on the left by $K$, and we apply to the quantity on the right the basic extension formula, we obtain:
$$K=\frac{1}{n-q-1}\left((\Sigma a_i+c+n-1)\Phi\begin{pmatrix}\ba&c\\ \bb&0\end{pmatrix}-\sum_{s=1}^q(a_s+1)\Phi\begin{pmatrix}\ba^{(s)}&c\\ \bb&0\end{pmatrix}-(c+1)K\right)$$

This gives the formula of $K$ in the statement.
\end{proof}

As a first consequence of our results, we can establish now a number of concrete formulae. The first such formula computes all the joint moments of $u_{11},u_{12},u_{21}$.

\begin{theorem}
We have the ``triangular formula''
$$\Phi\begin{pmatrix}a&c\\ b&0\end{pmatrix}=\frac{(n-1)!!}{(n-2)!!}\cdot\frac{(a+c+n-2)!!(b+c+n-2)!!}{(c+n-2)!!(a+b+c+n-1)!!}$$
valid for any $a,b,c\in 2\mathbb N$.
\end{theorem}

\begin{proof}
We prove this by induction over $c\in 2\mathbb N$. At $c=0$ this follows from the 1-row formula, so assume that this is true at $c$. By using Lemma 4.2, we get:
$$\Phi\begin{pmatrix}a&c+2\\ b&0\end{pmatrix}
=\frac{1}{n+c-1}\left((a+c+n-1)\Phi\begin{pmatrix}a&c\\ b&0\end{pmatrix}-(a+1)\Phi\begin{pmatrix}a+2&c\\ b&0\end{pmatrix}\right)$$

Let us call $L-R$ the above expression. According to the recurrence, we have:
$$L=\frac{(n-1)!!}{(n-2)!!}\cdot\frac{(a+c+n)!!(b+c+n-2)!!}{(c+n)!!(a+b+c+n-1)!!}$$
$$R=(a+1)\frac{(n-1)!!}{(n-2)!!}\cdot\frac{(a+c+n)!!(b+c+n-2)!!}{(c+n)!!(a+b+c+n+1)!!}$$

Thus we obtain the following formula:
\begin{eqnarray*}
\Phi\begin{pmatrix}a&c+2\\ b&0\end{pmatrix}
&=&\frac{(n-1)!!}{(n-2)!!}\cdot\frac{(a+c+n)!!(b+c+n-2)!!}{(c+n)!!(a+b+c+n+1)!!}((a+b+c+n)-(a+1))\\
&=&\frac{(n-1)!!}{(n-2)!!}\cdot\frac{(a+c+n)!!(b+c+n-2)!!}{(c+n)!!(a+b+c+n+1)!!}(b+c+n-1)\\
&=&\frac{(n-1)!!}{(n-2)!!}\cdot\frac{(a+c+n)!!(b+c+n)!!}{(c+n)!!(a+b+c+n+1)!!}
\end{eqnarray*}

Thus the formula to be proved is true at $c+2$, and we are done.
\end{proof}

As a first observation, by combining the above formula with the compression formula we obtain the following result, fully generalizing Theorem 1.2.

\begin{corollary}
We have the formula
$$\Phi\begin{pmatrix}a&c_1&\ldots&c_q\\ b&0&\ldots&0\end{pmatrix}=\frac{(n-1)!!}{(n-2)!!}\cdot\frac{(a+\Sigma c_i+n-2)!!(b+\Sigma c_i+n-2)!!}{(\Sigma c_i+n-2)!!(a+b+\Sigma c_i+n-1)!!}$$
valid for any even numbers $a,b$ and $c_1,\ldots,c_q$.
\end{corollary}

\begin{proof}
This follows indeed from Theorem 4.3 and from the compression principle.
\end{proof}

As a second observation, at $a=0$ the triangular formula computes all the joint moments of $u_{12},u_{21}$. To our knowledge, this quite basic result was previously unknown.

\begin{corollary}
The joint moments of $2$ orthogonal group coordinates $x,y\in\{u_{ij}\}$, chosen in generic position (i.e. not on the same row or column), are given by
$$\int_{O_n}x^\alpha y^\beta\,du=\frac{(n-2)!\alpha!!\beta!!(\alpha+\beta+n-2)!!}{(\alpha+n-2)!!(\beta+n-2)!!(\alpha+\beta+n-1)!!}$$
for $\alpha,\beta$ even, and vanish if one of $\alpha,\beta$ is odd.
\end{corollary}

\begin{proof}
By symmetry we may assume that our coordinates are $x=u_{12}$ and $y=u_{21}$, and the result follows from Theorem 4.3, with $a=0$, $c=\alpha$, $b=\beta$.
\end{proof}

\section{The flipping principle}

In this section we state and prove the main conceptual result in this paper. This is a non-trivial and quite powerful, but remarkably simple symmetry result, stating that $\Phi(^\ba_\bb)$ is invariant under the upside-down flipping of any column of $(^\ba_\bb)$.

Let us begin with the case of the elementary matrices.

\begin{lemma}
We have the formula
$$\Phi\begin{pmatrix}2^a&0^b\\ 0^a&2^b\end{pmatrix}=\frac{(n-1)!!}{(n-2)!!}\cdot\frac{(2a+2b+n-2)!!}{(2a+2b+n-1)!!}$$
valid for any $a,b\in\mathbb N$.
\end{lemma}

\begin{proof}
Indeed, by using the compression principle, we obtain:
$$\Phi\begin{pmatrix}2^a&0^b\\ 0^a&2^b\end{pmatrix}
=\Phi\begin{pmatrix}2a&0\\ 0&2b\end{pmatrix}\\
=\Phi\begin{pmatrix}0&2a\\ 2b&0\end{pmatrix}$$

On the other hand, by applying the triangular formula, we obtain:
$$\Phi\begin{pmatrix}0&2a\\ 2b&0\end{pmatrix}=\frac{(n-1)!!}{(n-2)!!}\cdot\frac{(2a+n-2)!!(2a+2b+n-2)!!}{(2a+n-2)!!(2a+2b+n-1)!!}$$

By simplifying the fraction, we obtain the formula in the statement.
\end{proof}

\begin{lemma}
We have the ``elementary flipping'' formula
$$\Phi\begin{pmatrix}1^{2s}&2^a&0^b\\ 1^{2s}&0^a&2^b\end{pmatrix}=\Phi\begin{pmatrix}1^{2s}&2^c&0^d\\ 1^{2s}&0^c&2^d\end{pmatrix}$$
valid for any $s\in\mathbb N$ and any $a,b,c,d\in\mathbb N$ satisfying $a+b=c+d$. 
\end{lemma}

\begin{proof}
We prove this result by induction over $s$. At $s=0$ this follows from the explicit formula in Lemma 5.1, because the right term there depends only on $a+b$.

So, assume that the result is true at $s\in\mathbb N$. We use the following equality, coming from the triangular formula:
$$\Phi\begin{pmatrix}2a&2c\\ 2b&0\end{pmatrix}=\Phi\begin{pmatrix}2a&0\\ 2b&2c\end{pmatrix}$$

Assume $a\geq b$ and consider the elementary expansion of the above two quantities, where $K_r(a,b)$ denotes the coefficient appearing in the elementary expansion formula:
$$\Phi\begin{pmatrix}2a&2c\\ 2b&0\end{pmatrix}=\sum_{r=0}^bK_r(a,b)\Phi\begin{pmatrix}1^{2r}&2^{a+c-r}&0^{b-r}\\ 1^{2r}&0^{a+c-r}&2^{b-r}\end{pmatrix}$$
$$\Phi\begin{pmatrix}2a&0\\ 2b&2c\end{pmatrix}=\sum_{r=0}^bK_r(a,b)\Phi\begin{pmatrix}1^{2r}&2^{a-r}&0^{b+c-r}\\ 1^{2r}&0^{a-r}&2^{b+c-r}\end{pmatrix}$$

We know that the sums on the right are equal, for any $a,b,c$ with $a\geq b$. With the choice $b=s$, this equality becomes:
$$\sum_{r=0}^sK_r(a,s)\Phi\begin{pmatrix}1^{2r}&2^{a+c-r}&0^{s-r}\\ 1^{2r}&0^{a+c-r}&2^{s-r}\end{pmatrix}=\sum_{r=0}^sK_r(a,s)\Phi\begin{pmatrix}1^{2r}&2^{a-r}&0^{s+c-r}\\ 1^{2r}&0^{a-r}&2^{s+c-r}\end{pmatrix}$$

Now by the induction assumption, the first $r$ terms of the above two sums coincide. So, the above equality tells us that the last terms ($r=s$) of the two sums are equal:
$$\Phi\begin{pmatrix}1^{2s}&2^{a+c-s}\\ 1^{2s}&0^{a+c-s}\end{pmatrix}=\Phi\begin{pmatrix}1^{2s}&2^{a-s}&0^c\\ 1^{2s}&0^{a-s}&2^c\end{pmatrix}$$

Since this equality holds for any $a\geq s$ and any $c$, this shows that the elementary flipping formula holds at $s$, and we are done.
\end{proof}

\begin{theorem}
We have the ``flipping formula''
$$\Phi\begin{pmatrix}\ba&\bc\\ \bb&\bd\end{pmatrix}=\Phi\begin{pmatrix}\ba&\bd\\ \bb&\bc\end{pmatrix}$$
valid for any vectors $\ba,\bb\in\mathbb N^p$ and $\bc,\bd\in\mathbb N^q$.
\end{theorem}

\begin{proof}
Consider indeed the elementary expansion of the two quantities in the statement, where $K_r(\ba,\bb)$ are the coefficients appearing in the elementary expansion formula:
$$\Phi\begin{pmatrix}2\ba&2\bc\\ 2\bb&2\bd\end{pmatrix}
=\sum_{r_is_j}\prod_{ij}K_{r_i}(a_i,b_i)K_{s_j}(c_j,d_j)\Phi\begin{pmatrix}1^{2R+2S}&2^{A+C-R-S}&0^{B+D-R-S}\\
1^{2R+2S}&0^{A+C-R-S}&2^{B+D-R-S}\end{pmatrix}$$
$$\Phi\begin{pmatrix}2\ba&2\bd\\ 2\bb&2\bc\end{pmatrix}
=\sum_{r_is_j}\prod_{ij}K_{r_i}(a_i,b_i)K_{s_j}(d_j,c_j)\Phi\begin{pmatrix}1^{2R+2S}&2^{A+D-R-S}&0^{B+C-R-S}\\
1^{2R+2S}&0^{A+D-R-S}&2^{B+C-R-S}\end{pmatrix}$$

Our claim is that two formulae are in fact identical. Indeed, the first remark is that the various indices vary in the same sets. Also, since the function $K_r(\ba,\bb)$ is symmetric in $\ba,\bb$, the numeric coefficients are the same. As for the $\Phi$ terms on the left, these are equal as well, due to elementary flipping formula, so we are done.
\end{proof}

Observe that in the case $p=q$, the flipping principle shows in fact that the function $\Phi(^\ba_\bb{\ }^\bc_\bd)$ is symmetric in its entries $\ba,\bb,\bc,\bd\in\mathbb N^p$. We will further develop this point of view in section 7 below, in the ``numeric'' case, $p=1$.

As a partial conclusion to the results in this paper, the two-row integrals obey to 3 general principles: compression, extension, and flipping. We will come back to these kind of questions at the end of the next section, with an extra invariance property.

\section{The two-row formula}

In this section we state and prove the main result in this paper: a concrete formula, in terms of sums of products of factorials, for the arbitrary two-row integrals.

We know from section 3 that these integrals are subject to an ``elementary expansion'' formula, so what is left to do is to compute the values of the elementary integrals.

These values are given by the following technical result.

\begin{lemma}
For any $a,b,r$ we have:
$$\Phi\begin{pmatrix}1^{2r}&2^a&0^b\\ 1^{2r}&0^a&2^b\end{pmatrix}=(-1)^r\frac{(n-1)!!}{(n-2)!!}\cdot\frac{(2r)!!(2a+2b+2r+n-2)!!}{(2a+2b+4r+n-1)!!}$$
\end{lemma}

\begin{proof}
As a first observation, at $r=0$ the result follows from Lemma 5.1.

Consider the elementary expansion formula, with $a,b\in\mathbb N$, $a\geq b$:
$$\Phi\begin{pmatrix}2a\\ 2b\end{pmatrix}=\sum_{r=0}^b\frac{4^ra!b!}{(2r)!(a-r)!(b-r)!}\Phi\begin{pmatrix}1^{2r}&2^{a-r}&0^{b-r}\\
1^{2r}&0^{a-r}&2^{b-r}
\end{pmatrix}$$

By using the ``flipping principle'', this formula becomes:
$$\Phi\begin{pmatrix}2a\\ 2b\end{pmatrix}=\sum_{r=0}^b\frac{4^ra!b!}{(2r)!(a-r)!(b-r)!}\Phi\begin{pmatrix}1^{2r}&2^{a+b-2r}\\
1^{2r}&0^{a+b-2r}
\end{pmatrix}$$

The point is that the quantity on the left is known, and this allows the computation of the integrals on the right. More precisely, let us introduce the following function:
$$\varphi_r(a)=\Phi\begin{pmatrix}1^{2r}&2^a\\
1^{2r}&0^a
\end{pmatrix}$$

Then the above equality translates into the following equation:
$$\Phi\begin{pmatrix}2a\\ 2b\end{pmatrix}=\sum_{r=0}^b\frac{4^ra!b!}{(2r)!(a-r)!(b-r)!}\varphi_r(a+b-2r)$$

According to Theorem 1.2 and Proposition 3.2, the values on the left are given by:
$$\Phi\begin{pmatrix}2a\\ 2b\end{pmatrix}=\frac{(n-1)!!(2a+n-2)!!(2b+n-2)!!}{(n-2)!!(n-2)!!(2a+2b+n-1)!!}$$

Now by taking $b=0,1,2,\ldots$, the above equations will succesively produce the values of $\varphi_r(a)$ for $r=0,1,2,\ldots$, so we have here an algorithm for computing these values.

On the other hand, a direct computation based on standard summation formulae shows that our system is solved by the values of $\varphi_r(a)$ given in the statement, namely:
$$\varphi_r(a)=(-1)^r\frac{(n-1)!!}{(n-2)!!}\cdot\frac{(2r)!!(2a+2r+n-2)!!}{(2a+4r+n-1)!!}$$

Now by using one more time the ``flipping principle'', the knowledge of the quantities $\varphi_r(a)$ fully recovers the general formula in the statement, and we are done.
\end{proof}

We are now in position of stating and proving the main result in this paper.

\begin{theorem}
The $2$-row integrals are given by the formula
$$\Phi\begin{pmatrix}2\ba\\ 2\bb\end{pmatrix}=\frac{(n-1)!!}{(n-2)!!}\sum_{r_1,\ldots,r_q}(-1)^R\prod_{i=1}^q\frac{4^{r_i}a_i!b_i!}{(2r_i)!(a_i-r_i)!(b_i-r_i)!}\cdot\frac{(2R)!!(2S-2R+n-2)!!}{(2S+n-1)!!}$$
where the sum is over $r_i=0,1,\ldots,\min(a_i,b_i)$, and $S=\Sigma a_i+\Sigma b_i,R=\Sigma r_i$.
\end{theorem}

\begin{proof}
This follows from the elementary expansion formula, by plugging in the explicit values for the elementary integrals, that we found in Lemma 6.1.
\end{proof}

As a first remark, all the results in the previous sections, and in particular the compression, extension, and flipping principles, can be deduced from the above formula. 

We would like to present now another invariance principle, which is somehow of different nature, because it involves a ``transmutation'' of the $n$ variable. The strange nature of this principle comes as well from the fact that is not clear how to obtain it directly.

\begin{lemma}
We have the ``basic transmutation'' formula
$$\Phi_n\begin{pmatrix}\ba&2\\ \bb&0\end{pmatrix}=\frac{n-1}{n}\,\Phi_{n+2}\begin{pmatrix}\ba\\ \bb\end{pmatrix}$$
valid for any two vectors $\ba,\bb\in(2\mathbb N)^q$.
\end{lemma}

\begin{proof}
It is convenient to replace $\ba,\bb$ by their doubles $2\ba,2\bb$. Consider the elementary expansion of our two quantities, where $K$ denotes as usual the numeric coefficients:
$$\Phi_n\begin{pmatrix}2\ba&2\\ 2\bb&0\end{pmatrix}=\sum_{r_1\ldots r_q}\prod_iK_{r_i}(a_i,b_i)\Phi_n\begin{pmatrix}1^{2R}&2^{A-R+1}&0^{B-R}\\ 1^{2R}&0^{A-R+1}&2^{B-R}\end{pmatrix}$$
$$\Phi_{n+2}\begin{pmatrix}2\ba\\ 2\bb\end{pmatrix}=\sum_{r_1\ldots r_q}\prod_iK_{r_i}(a_i,b_i)\Phi_{n+2}\begin{pmatrix}1^{2R}&2^{A-R}&0^{B-R}\\ 1^{2R}&0^{A-R}&2^{B-R}\end{pmatrix}$$

In these two formulae the sums are over the same indices, and the numeric coefficients are the same. So, it is enough to prove that we have the following equality:
$$\Phi_n\begin{pmatrix}1^{2R}&2^{A-R+1}&0^{B-R}\\ 1^{2R}&0^{A-R+1}&2^{B-R}\end{pmatrix}=\frac{n-1}{n}\Phi_{n+2}\begin{pmatrix}1^{2R}&2^{A-R}&0^{B-R}\\ 1^{2R}&0^{A-R}&2^{B-R}\end{pmatrix}$$

That is, we just have to prove the basic transmutation formula for the elementary matrices. But this follows from the explicit formula in Lemma 6.1, and we are done.
\end{proof}

\begin{theorem}
We have the ``transmutation formula''
$$\Phi_n\begin{pmatrix}\ba&\bc\\ \bb&0\end{pmatrix}=\frac{(n-1)!!}{(n-2)!!}\cdot\frac{(\Sigma c_i+n-2)!!}{(\Sigma c_i+n-1)!!}\,\Phi_{n+\Sigma c_i}\begin{pmatrix}\ba\\ \bb\end{pmatrix}$$
valid for any vectors $\ba,\bb\in(2\mathbb N)^q$ and $\bc\in(2\mathbb N)^p$.
\end{theorem}

\begin{proof}
First, by using the compression principle, we just have to prove the above formula at $p=1$. That is, we have to prove the following formula:
$$\Phi_n\begin{pmatrix}\ba&c\\ \bb&0\end{pmatrix}=\frac{(n-1)!!}{(n-2)!!}\cdot\frac{(c+n-2)!!}{(c+n-1)!!}\,\Phi_{n+c}\begin{pmatrix}\ba\\ \bb\end{pmatrix}$$

But this follows from the basic transmutation formula. Indeed, by applying this formula $c$ times, and by using the compression principle, we obtain:
\begin{eqnarray*}
\Phi_n\begin{pmatrix}\ba&2c\\ \bb&0\end{pmatrix}
&=&\frac{n-1}{n}\cdot\frac{n+1}{n+2}\ldots\frac{n+2c-3}{n+2c-2}
\,\Phi_{n+2c}\begin{pmatrix}\ba\\ \bb\end{pmatrix}\\
&=&\frac{(n+2c-2)!!}{(n-2)!!}\cdot\frac{(n-1)!!}{(n+2c-1)!!}\,
\Phi_{n+2c}\begin{pmatrix}\ba\\ \bb\end{pmatrix}\\
&=&\frac{(n-1)!!}{(n-2)!!}\cdot\frac{(n+2c-2)!!}{(n+2c-1)!!}\,
\Phi_{n+2c}\begin{pmatrix}\ba\\ \bb\end{pmatrix}
\end{eqnarray*}

This gives the formula in the statement.
\end{proof}

\section{The 2 x 2 case}

In this section we discuss the case of the $2\times 2$ matrices $a\in M_2(\mathbb N)$, with a number of refinements of the results in the previous sections. As we will see right away, there is some ``magic'' in the $2\times 2$ case, waiting to be fully discovered, and conceptually understood. The present section should be rather regarded as an introduction to the $2\times 2$ problematics.

The ``magic'' comes from the following result.

\begin{theorem}
The function
$$f(a,b,c,d)=\frac{I\begin{pmatrix}a&c\\ b&d\end{pmatrix}}{(a+d+n-2)!!(b+c+n-2)!!}$$
is symmetric in $a,b,c,d$.
\end{theorem}

\begin{proof}
This follows from the flipping formula for the $2\times 2$ matrices. Indeed, by using the conversion formula in Proposition 3.2, we obtain:
$$f(a,b,c,d)=\frac{(n-2)!!(n-2)!!a!!b!!c!!d!!\Phi\begin{pmatrix}a&c\\ b&d\end{pmatrix}}{(a+d+n-2)!!(b+c+n-2)!!(a+c+n-2)!!(b+d+n-2)!!}$$

The flipping principle tells us that the $\Phi$ quantity is symmetric in $c,d$. Now since the coefficient is symmetric as well in $c,d$, we conclude that $f$ is symmetric in $c,d$. Together with the standard fact that $f$ is symmetric in $a,d$, and also in $b,c$, this gives the result.
\end{proof}

The problem now is to find a formula for $f(a,b,c,d)$, as a ``sum of products of factorials, symmetric in $a,b,c,d$''. Let us introduce the following basic quantities.

\begin{definition}
Associated to $a,b,c,d$ are the following quantities:
\begin{enumerate}
\item $S_k(a,b,c,d)=(a+k)(b+k)(c+k)(d+k)$.

\item $F_k(a,b,c,d)=(a+k)!!(b+k)!!(c+k)!!(d+k)!!$.

\item $P_k(a,b,c,d)=(a+b+c+d+k)!!$.
\end{enumerate}
\end{definition}

Observe that all the above quantities are symmetric in $a,b,c,d$.

\begin{theorem}
We have the following formula:
$$f(a,b,c,0)=\frac{(n-2)!(n-2)!!}{P_{n-1}}\cdot\frac{F_0}{F_{n-2}}$$
\end{theorem} 

\begin{proof}
From Theorem 4.3 we get:
$$\Phi\begin{pmatrix}a&c\\ b&0\end{pmatrix}=\frac{(n-1)!!}{(n-2)!!}\cdot\frac{(a+c+n-2)!!(b+c+n-2)!!}{(c+n-2)!!(a+b+c+n-1)!!}$$

By using the above conversion formula between $f$ and $\Phi$, we obtain:
$$f(a,b,c,0)=\frac{(n-2)!a!!b!!c!!}{(a+n-2)!!(b+n-2)!!(c+n-2)!!(a+b+c+n-1)!!}$$

This gives the formula in the statement.
\end{proof}

In the general case, Theorem 6.2 provides of course a concrete formula for $f(a,b,c,d)$, as a double sum of products of factorials. However, since that formula is not obviously symmetric in $a,b,c,d$, there is definitely room here for some improvements. 

Instead of getting into this subject, which is a bit away from the purposes of this paper, let us just write down the final formula in the $2\times 2$ case, in the form of a conjecture.

\begin{conjecture}
For $a,b,c,d$ even we have the formula
$$f(a,b,c,d)=\frac{(n-2)!(n-2)!!}{P_{n-1}}\cdot\frac{F_0}{F_{n-2}}\sum_{r=0}^\infty\frac{n+4r-3}{n-3}\begin{pmatrix}n+2r-4 \\ 2r \end{pmatrix}\frac{S_0S_{-2}\ldots S_{-2r+2}}{S_{n-1}S_{n+1}\ldots S_{n+2r-3}}$$
where the sum is actually finite, stopping at $l=\min(a,b,c,d)/2$.
\end{conjecture}

As already mentioned, this formula should follow from Theorem 6.2, by doing some summation work. While the techniques here, such as Sister Celine's method, don't lack (see \cite{pwz}), we would rather keep this technical work for one of our future papers.

Let us also mention that the above formula was actually obtained at an early stage of the present work, and comes with heavy computer evidence.

Finally, in the case where $a,b,c,d$ are odd, the situation is quite similar. An analogue of Theorem 7.1 holds, in the sense that the function $f$ given by the formula there can be shown to be symmetric in $a,b,c,d$. We have the following conjectural formula:

\begin{conjecture}
For $a,b,c,d$ odd we have the formula
$$f(a,b,c,d)=-\frac{(n-2)! n!!}{P_{n-1}}\cdot\frac{F_1}{F_{n-1}}\sum_{r=0}^\infty\frac{n+4r-1}{(n-1)(n-3)}\begin{pmatrix}n+2r-3 \\ 2r+1 \end{pmatrix}\frac{S_{-1}S_{-3}\ldots S_{-2r+1}}{S_nS_{n+2}\ldots S_{n+2r-2}}$$
where the sum is actually finite, stopping at $l=(\min(a,b,c,d)-1)/2$.
\end{conjecture}

Once again, this is a statement coming with heavy computer evidence. As regarding a potential proof, this should come from a suitable extension of Theorem 6.2 to the ``odd'' case. But getting into this technical subject is beyond the purposes of this paper.  

As a last remark, it is not clear how to unify the above ``odd'' formula with the ``even'' one from Conjecture 7.4. A more elaborated definition for the quantities $S,F,P$, and perhaps even for the double factorials themselves, is probably needed here.

\section{Concluding remarks}

We have seen in this paper that the two-row integrals of type $I(\ba)$ enjoy remarkable symmetry properties, which can be effectively used for their exact computation.

It is our hope that the present results will substantially contribute to the further development of the general study of integrals of type $I(\ba)$. As explained in \cite{bcs}, the question of computing exactly these integrals, which perhaps lacks a bit of motivation in the general context of ``probabilistic'' mathematical physics, where the $n\to\infty$ limit is usually the correct quantity to look at, is however a very interesting one in the context of pure mathematics, due to its relation with the Hadamard conjecture. So, our hope is that the present work will be a useful adding to the lineup of recent papers \cite{bcs}, \cite{cma}, \cite{mno}, \cite{ban}, \cite{zin}.

Finally, let us mention that there are some interesting ``noncommutative analogues'' of the problems investigated in the present paper. The free analogue of the hyperspherical law was recently found in \cite{bcz}, and some further occurences of this law come from \cite{bgo}, \cite{cur}. However, many questions are still open. For instance computing the ``exact correlation'' between 2 coordinates following this law remains a remarkably difficult, open problem.

\end{document}